\documentclass[iop]{emulateapj}
\usepackage{apjfonts}
\usepackage{epsfig}
\usepackage{amsmath}

\usepackage{natbib}
\usepackage[backref,breaklinks,colorlinks,citecolor=blue]{hyperref}        

\usepackage[all]{hypcap}

\def\chandra{\emph{Chandra}}
\def\as{$^{\prime\prime}$}
\def\msun{$M_{\odot}$}

\begin{document} 

\title{The merging galaxy cluster A520 --- a broken-up cool core, a dark
  subcluster, and an X-ray channel}

\shorttitle{The merging galaxy cluster A520}

\author{Qian H. S. Wang\altaffilmark{1}, Maxim Markevitch\altaffilmark{2,3}, Simona
Giacintucci\altaffilmark{1,3}}

\shortauthors{Wang, Markevitch \& Giacintucci}

\altaffiltext{1}{Department of Astronomy, University of Maryland, College
  Park, MD 20742}

\altaffiltext{2}{Astrophysics Science Division, NASA Goddard Space Flight
  Center, Greenbelt, MD 20771}

\altaffiltext{3}{Joint Space-Science Institute, University of Maryland,
  College Park, MD, 20742}

\begin{abstract}

We present results from a deep \chandra\ X-ray observation of a merging galaxy
cluster A520. A high-resolution gas temperature map reveals a long trail of
dense, cool clumps --- apparently the fragments of a cool core that has been
stripped from the infalling subcluster by ram pressure. The clumps should
still be connected by the stretched magnetic field lines. The observed
temperature variations imply that thermal conductivity is suppressed by a
factor $>100$ across the presumed direction of the magnetic field (as found in
other clusters), and is also suppressed {\em along}\/ the field lines by a
factor of several. Two massive clumps in the periphery of A520, visible in the
weak lensing mass map and the X-ray image, have apparently been completely
stripped of gas during the merger, but then re-accreted the surrounding
high-entropy gas upon exit from the cluster. The mass clump that hosted the
stripped cool core is also reaccreting hotter gas. An X-ray hydrostatic mass
estimate for the clump that has the simplest geometry agrees with the lensing
mass. Its current gas mass to total mass ratio is very low, 1.5--3\%, which
makes it a ``dark subcluster''.  We also found a curious low X-ray brightness
channel (likely a low-density sheet in projection) going across the cluster
along the direction of an apparent secondary merger. The channel may be caused
by plasma depletion in a region of an amplified magnetic field (with plasma
$\beta\sim 10-20$). The shock in A520 will be studied in a separate paper.

\end{abstract}


\section{Introduction}

Galaxy clusters form and grow via mergers of less massive systems in a
hierarchical process governed by gravity \citep[e.g.,][]{Press74, Springel06}.
In the course of each merger, approximately speaking, the kinetic energy
carried by the gas of the colliding clusters dissipates into thermal energy
via shocks and turbulence and, in the absence of further disturbances, the
hotter gas comes into approximate hydrostatic equilibrium with the deeper
gravitational potential of the resulting bigger cluster (e.g.,
\citealt{Bahcall77}) on a $\sim$Gyr timescale. What happens during that
Gigayear of violent gas motions is very interesting, because it can illuminate
several aspects of the physics of the intracluster plasma
\citep[e.g.,][]{2007PhR...443....1M}. Ram pressure of the gas flows may
strip the subclusters of their gas \citep[e.g.,][]{2006ApJ...648L.109C} and
disturb and even destroy their cool cores either directly
\citep[e.g.,][]{1991MNRAS.252P..17F, 2000ApJ...541..542M} or by facilitating
mixing with the surrounding gas \citep{2010ApJ...717..908Z}. Temperature
gradients in the gas generated by shock heating and mixing of different gas
phases should be quickly erased by thermal conduction, if it is not
suppressed \citep[e.g.,][]{2003ApJ...586L..19M, 2012A&A...541A..57E}. All of
this makes observations of merging clusters in the X-ray, where we can map
the density and temperature of the hot intracluster plasma, extremely
interesting.

\begin{figure*}
        \centering
        \leavevmode
        \includegraphics*[width=180mm]{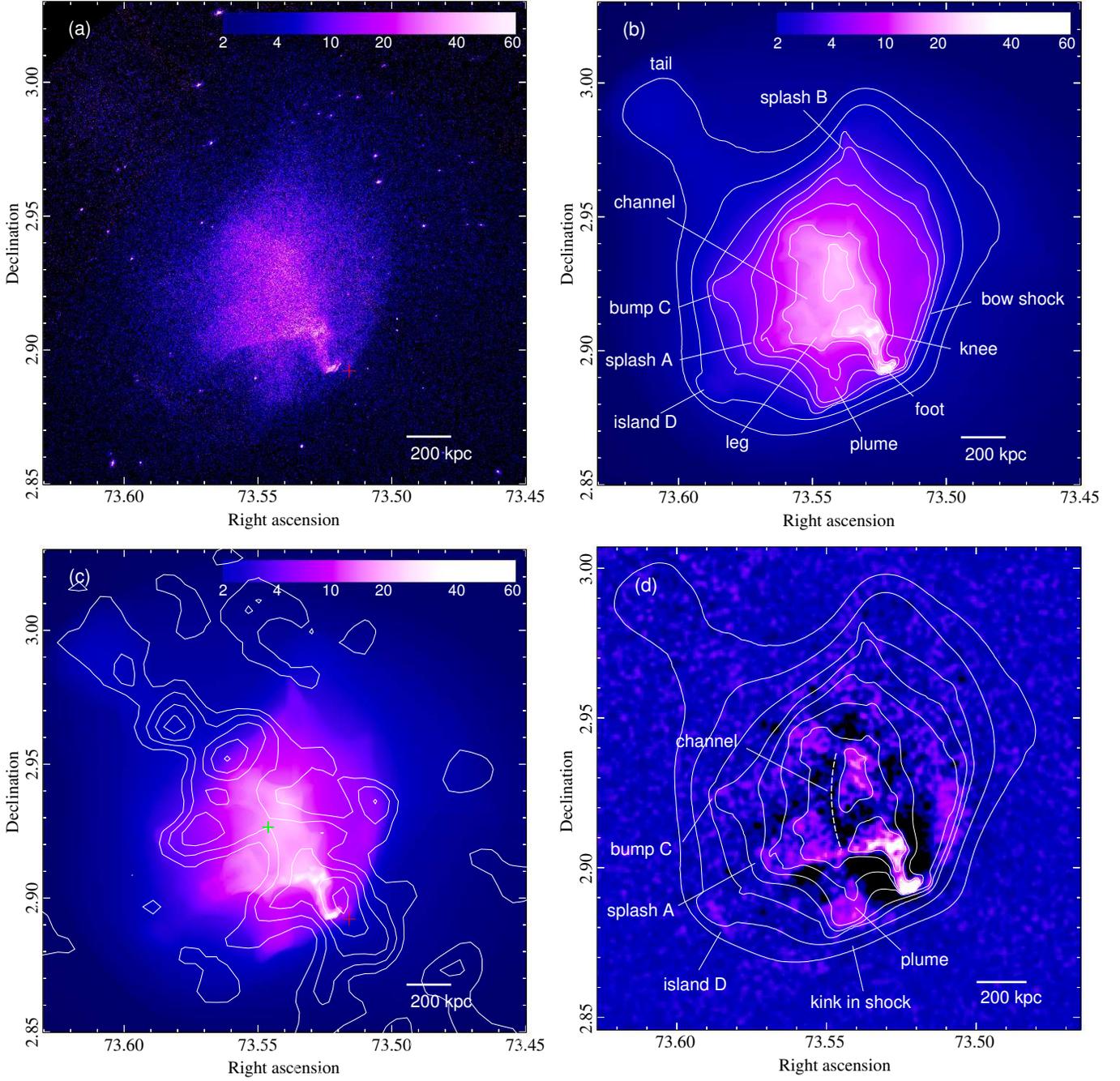}

        \caption{ ({\em a}) \chandra\ 0.8--4~keV surface brightness binned to
          1\as\ pixels, without smoothing or source removal. The color scale
          is in units of $10^{-6}$~counts~s$^{-1}$~arcsec$^{-2}$. The red
          cross marks the center of the BCG, offset from the bright tip by
          about 20\as\ = 67~kpc. ({\em b}) Wavelet smoothing of the image in
          panel {\em a}, with point sources removed, with brightness
          contours spaced by factor 1.5. ({\em c}) Weak lensing mass
          contours from D.\ Clowe (private communication), overlaid on the
          wavelet X-ray image from panel {\em b}. The contour levels (mass
          surface density, linear step) are same as in
          \cite{2012ApJ...758..128C}. Green cross marks the position of the
          contested ``dark clump'', the red cross marks the BCG.  ({\em d})
          Residual X-ray image after subtracting the $>210$ kpc scale
          wavelets components, slightly smoothed. Prominent features
          discussed in this paper are marked in panels {\em b}\/ and {\em
            d}.}

        \label{fig:overview}
\end{figure*}

The hot ($T\simeq 7$~keV, \citealt{2004ApJ...605..695G}) galaxy cluster Abell
520 at $z=0.203$ \citep{1975ApJ...197L..95W} is one of only a handful of
merging systems with a shock front clearly visible in the sky plane
\citep{2005ApJ...627..733M}, which makes the merger geometry quite
unambiguous. The cluster has a detailed map of the projected total mass
distribution derived from weak gravitational lensing data
\citep{2007ApJ...668..806M, 2008PASJ...60..345O, 2012ApJ...747...96J,
2012ApJ...758..128C, 2014ApJ...783...78J}. We show an uncropped version of the
mass map from \cite{2012ApJ...758..128C}, provided by D.\ Clowe (private
communication), in \autoref{fig:overview}{\em c}. While the above authors
disagree on the details (in particular, \citeauthor{2007ApJ...668..806M} and
\citeauthor{2014ApJ...783...78J} reported the presence of a ``dark clump''
with an anomalously high $M/L$ ratio in the middle of the cluster, marked by a
green cross in \autoref{fig:overview}{\em c}, while
\citeauthor{2012ApJ...758..128C} contested its statistical significance), the
lensing maps agree qualitatively quite well. The overall picture is a ``train
wreck'' of several mass clumps mostly aligned in a chain along the NE-SW
direction.  This is consistent with the merger direction indicated by the
X-ray shock front.

In this paper, we analyze in detail an extra-deep 0.5~Ms \chandra\ observation
of A520. It will allow us to gain insights into many of the above physical
processes, such as the cool core stripping and the suppression of thermal
conductivity. Analysis of the shock front based on the same X-ray data,
supplemented by the archival radio data, will be given in a future paper
(Wang, Giacintucci, \& Markevitch 2016, in prep.).

We assume a flat cosmology with $H_0=70$ km s$^{-1}$ Mpc$^{-1}$ and
$\Omega_m=0.3$, in which 1\as~ is 3.34 kpc at $z=0.203$. Errors are quoted
at 90\% confidence in text, and at 1-$\sigma$ in figures, unless otherwise
stated.

%%%% 3.338 kpc/"; 1 px = 1.6425 kpc

%%% FIGURE 1

\section{X-ray data analysis}
\label{sec:xray}

We analyzed observations of A520 with \chandra\ Advanced CCD Imaging
Spectrometer (ACIS) between December 2007 and January 2008 for a total of 447
ks (ObsIDs 9424, 9425, 9426, 9430). This gave 423 ks after cleaning for flares
as described in the next paragraph. We chose not to combine these with earlier
observations (ObsIDs 528, 4215, and 7703 with exposure times 9.47 ks, 66.27
ks, and 5.08 ks, respectively). The two short observations will not
meaningfully improve our results so we omitted them for simplicity. ObsID 4215
is affected by a long low-level background flare, which
\cite{2005ApJ...627..733M} modelled as an excess over the quiescent background
and propagated the error for spectral modelling. Seeing this would increase
our total exposure time by at most 15\%, yet potentially introduce more
uncertainty to background subtraction (see \autoref{sec:bg}) we chose not to
complicate our subsequent analysis.

We reprocessed Level=1 event files using \texttt{acis\_process\_events} of the
\chandra\ X-ray Center (CXC) software, CIAO (4.6).%
\footnote{http://cxc.harvard.edu/ciao}
We applied the standard event filtering procedure of masking bad pixels, grade
filtering, removal of cosmic ray afterglow and streak events and the detector
background events identified using the VFAINT mode data.  Periods of elevated
background were identified using the 2.5--7~keV light curve in a background
region free of cluster emission on the ACIS-I chips (by excluding a circle of
$r=7'$ centered on A520 and another circle of $r=1.5'$ on a small extended
source to the SW). Time bins of 1 ks were used, and bins whose count rates
were more than 20\% different from the mean value were discarded, resulting in
423 ks of total clean exposure. During the clean exposure, no gradual changes
in the quiescent background level were apparent during any of the
observations; the mean rates varied with time by less than 10\%. We also
checked that there was no time variability in the ratio of the 2.5--7~keV to
9.5--12~keV counts using time bins of 10 ks. The mean value of this ratio was
also in good agreement (within 2\%) of that in the blank-sky background
dataset.  The latter two checks ensure the absence of faint residual
background flares and the accuracy of modeling the detector background using
the blank-sky dataset \citep{2006ApJ...645...95H} that we describe below.

The ACIS readout artifact was modeled using
\texttt{make\_readout\_bg}%
\footnote{http://cxc.harvard.edu/contrib/maxim/make\_readout\_bg}
and treated as an additional background component in our analysis (as in
\citealt{2000ApJ...541..542M}).

To create flux images, exposure maps were created using
Alexey Vikhlinin's tools.%
\footnote{http://hea-www.harvard.edu/$\sim$alexey/CHAV}
The exposure maps account for the position- and energy-dependent variation in
effective area and detector efficiency. The exposure maps for different
observations were co-added in sky coordinates. Then, the co-added
background-subtracted counts images were divided by the total exposure map to
produce a flux image. The four observations of A520 were set up with small
relative offsets in the sky to minimize the effect of chip gaps on the final
total image. 

We excluded point sources from our analysis by visually inspecting the
0.8--4~keV and 2--7~keV images at different image binning and smoothing
scales.

\subsection{Sky background}
\label{sec:bg}

To model the detector and sky background, we used the ACIS blank-sky
background dataset from the corresponding epoch (``period E'') as described in
\cite{2003ApJ...586L..19M} and \cite{2006ApJ...645...95H}. The VFAINT mode
filter was applied; the events were projected to the sky for each observation
using \texttt{make\_acisbg}.%
\footnote{http://cxc.harvard.edu/contrib/maxim/acisbg/}
The count rate derived from the background data was then scaled so that it had
the same 9.5--12~keV counts as the observed data. This was further reduced by
1.32\% to accommodate the amount of background contained in the readout
artifact. For flux images, this was done by multiplying the background counts
image by a rescaling factor. For spectral analysis, this was effected by
setting the BACKSCAL keyword in the spectra FITS files.

After subtracting the ACIS background normalized by the 9.5--12~keV rate, the
90\% uncertainty of the 0.8--9~keV quiescent backgound normalization is 3\%
\citep{2006ApJ...645...95H}. We will vary the background normalization by this
amount to estimate its contribution to the overall uncertainties. This is
particularly important for the low surface brightness cluster outskirt for
which the temperature uncertainties are dominated by the background; hence our
decision to exclude ObsID 4215 in our analysis due to the presence of a flare.

After subtracting the blank-sky and readout artifact backgrounds, the spectrum
of the cluster-free background region revealed a small positive residual flux
mostly at $E\sim 2$~keV. Some residual (positive or negative) is expected, as
the soft CXB varies across the sky and the blank-sky dataset comes from other
regions of the sky. We modeled this residual in the 0.5--9~keV band with an
empirical spectral model consisting of two APEC components, a power law and a
Gaussian. The thermal components were set to temperatures of 0.2~keV and
0.4~keV and their normalizations were allowed to vary, based on the study of
the soft CXB \citep{2003ApJ...583...70M}. The Gaussian component best fit was
at $E=0.92 \pm 0.02$ keV with zero width ($\sigma < 0.04$~keV). The power law
component was added to account for the residuals above 2~keV, and it was found
that a photon index of 0.6 made a qualitative improvement. Of course, there is
no physical significance to this empirical model, as it describes a difference
between the true CXB (and possibly a very faint residual flare emission) and
the CXB components included in the blank-sky dataset. An alternative is to use
the ``stowed'' ACIS background dataset, which contains only the detector
background, and add physically-motivated CXB components. However, the
available stowed background dataset has a much shorter exposure than the
blank-sky dataset, which is critically important for our extra-deep A520
observation.  We assumed that our empirical residual background was constant
across the FOV (before the telescope vignetting), and included this model,
adjusted for sky area and exposure time, when doing spectral fits for the
cluster regions. For the narrow-band flux images, the residual was accounted
for by subtracting a constant value such that the flux in the background
region was zero. A520 is sufficiently small and there is enough cluster-free
area within the FOV to make this additional background modeling step possible.

\subsection{Spectral analysis}

The instrument responses for spectral analysis were generated as described
in \cite{2005ApJ...628..655V}. We used the CHAV tool \texttt{runextrspec} to
generate the PHA, ARF and RMF files for each pointings. The PHA files
(observed data, blank-sky background and readout background) were co-added
using \texttt{addspec} from FTOOLS package. The \texttt{addarf} and
\texttt{addrmf} from FTOOLS were used to add ARFs and RMFs. They were
weighed by 0.5--2~keV counts in the applicable spectral extraction region.
 
Spectral analysis was performed in XSPEC (version 12.8.2). A
single-temperature fit to the cluster in a $3'$ circle (0.6 Mpc) centered on
soft band flux centroid at ($\alpha,\delta$)=(04:54:09.7, +02:55:25) (FK5,
J2000) gives $T=8.3 \pm 0.3$~keV, metal abundance $0.21 \pm 0.02$
\citep[relative to][]{1989GeCoA..53..197A} and absorption column $N_{\rm H} =
(6.3\pm 0.7)\times10^{20}~\mathrm{cm}^{-2}$. Factored into the error are
formal error from fitting, effect of the modeled soft residual background and
the 3\% uncertainty of the blank-sky background (\S\ref{sec:bg}); these were
added in quadrature. We fitted all spectra in the 0.8--9~keV band, excluding
the 1.7--1.9~keV and 7.3--7.6~keV intervals that are occasionally affected by
the detector features. The best-fit Galactic $N_{\rm H}$ is consistent with
$5.7\times10^{20}~\mathrm{cm}^{-2}$ from the LAB survey
\citep{2005A&A...440..775K}; with $N_{\rm H}$ fixed at the LAB value, we
obtain $T=8.5\pm0.3$~keV, while abundance is the same. It is also in good
agreement with the HI+H2 column density of $6.9\times 10^{20}\text{ cm}^{-2}$
\citep{2013MNRAS.431..394W} \footnote{Online tool:
http://www.swift.ac.uk/analysis/nhtot/index.php}. In subsequent
spatially-resolved analysis, we chose to use our best-fit value of $n_{\rm H}$
in order to compensate for any inaccuracies in the calibration of the
time-dependent ACIS low-energy response (while our choice of the 0.8~keV lower
energy cutoff should minimize their effect). We also fixed the abundance at
its best-fit value, as many of our fitting regions do not have enough counts
to constrain either $N_{\rm H}$ or abundance.

\section{Temperature maps}
\label{sec:tmap}

\begin{figure*}
        \centering
        \leavevmode
        \includegraphics*[width=180mm]{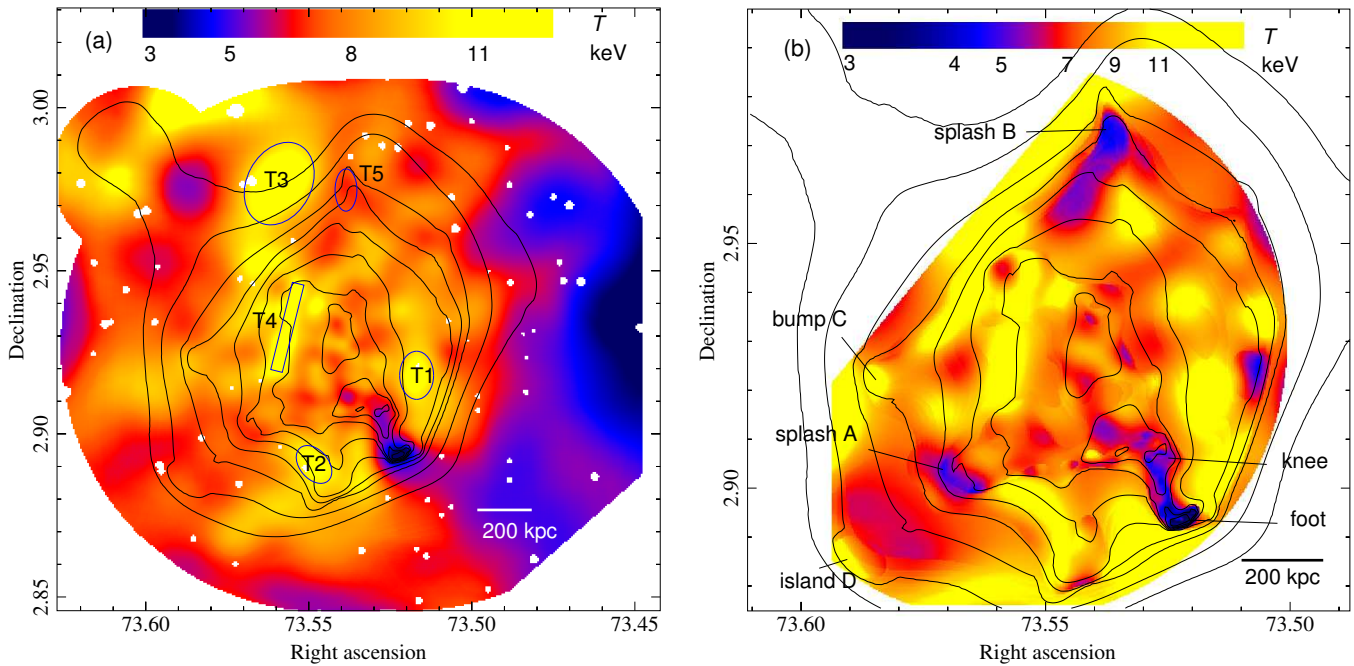}

        \caption{ ({\em a}) Variable smoothing temperature map with X-ray
          contours. The holes in the map are masked point sources. ({\em b})
          Wavelet temperature map.  The fitted images were reconstructed from
          component wavelet scales of 6.6, 13, 26, 53, 105 and 210 kpc.}

        \label{fig:tmap}
\end{figure*}

Temperature maps shown in \autoref{fig:tmap} were derived following the
method described in \cite{2000ApJ...541..542M}. We extracted 6 narrow band
flux images between 0.8--9~keV, excluding the 1.7--1.9~keV edge and 7.3--7.6
keV (possibly affected by poor subtraction of the instrumental lines). Both
flux and error images were smoothed prior to deriving the temperature map. A
single temperature MEKAL model was then fitted to a set of 6 flux values
from each pixel of the image, resulting in a smoothed temperature map. The
absorption column and metal abundance were fixed to the cluster best fit
values. Two smoothing methods were used, as described below.

%%% FIGURE 2

\subsection{Smoothing with variable-width Gaussian kernel}
\label{sec:gausstmap}

For this approach, the narrow-band images were smoothed using a Gaussian
kernel whose width at each image pixel is determined by surface brightness in
the 0.8--4~keV band, with the goal of preserving detail in bright regions. As
a reference for this smoothing method --- but also as a high-quality X-ray
image that shows the cluster structure on all scales and omits point sources
--- we used wavelet reconstruction of the 0.8--4~keV image, as described in
\cite{1998ApJ...502..558V}. We derived it using Alexey Vikhlinin's
\texttt{wvdecomp} tool in ZHTOOLS.%
\footnote{http://hea-www.harvard.edu/RD/zhtools/}
We extracted wavelet components (with the {\em atrous}\/ kernel and scales
increasing in geometric progression) on scales of 53, 105, 210 and 420 kpc (or
15.7\as, 31.5\as, 63.0\as and 126\as). Point sources are contained in wavelet
components on smaller scales than the first scale above, thus not included.
These image components were then co-added with the residual image smoothed by
the 840 kpc scale wavelet. This procedure retains most of the
statistically-significant extended structures on various angular scales. The
resulting image is shown in \autoref{fig:overview}{\em b}\/ next to the
original unsmoothed image; we will use it as reference when discussing various
features in this cluster.

Based on this reference image, the narrow-band images and their corresponding
error images (with point sources excised) were identically smoothed by a
variable-width Gaussian. By inspecting the error in the derived temperature
map, we determined that using the Gaussian smoothing radius $\propto$
flux$^{-0.7}$, and smoothing radius between 13 kpc and 200 kpc, achieved a
balance between revealing the temperature variations and suppressing noise.
The resulting temperature map is shown in \autoref{fig:tmap}{\em a}.

To check the validity of values shown in the map, we extracted the proper
spectra in a few hot spots $>10$~keV and a cooler spot (regions T1-T4 and T5
in \autoref{fig:tmap}{\em a}, respectively) and fitted their projected
temperatures in XSPEC. For T1, we obtain $12.1_{-2.4}^{+3.4}$~keV; for T2,
$11.3_{-2.6}^{+4.1}$ keV; for T3, we could only obtain a lower bound of
11.9~keV. For T4, the fit is $12.2_{-1.9}^{+2.5}$~keV, and for T5,
$6.4_{-1.4}^{+2.2}$~keV --- all values close to those in the smoothed map.

\subsection{Wavelet-smoothed temperature map}
\label{sec:wavtmap}

The second method uses wavelet image decomposition to identify structures at
different angular scales, and leave only the wavelet components on the
scales of interest in the narrow-band images used for temperature fitting,
instead of simple Gaussian smoothing. This method allows us, for example, to
subtract the structures on large angular scales and recover the temperature
contrast of features on the interesting small scales by reducing the
projection effects. Of course, such ``deprojection'' can only be
qualitative, as we do not know the gas distribution along the l.o.s.
and have to assume that structures on different scales are simply projected.
Nevertheless, for the interesting high-contrast features in A520, this
assumption should be close to reality.

This method has the greatest utility to recover the small-scale, cool, bright
structure at the ``foot'' and ``leg'' of A520 (\autoref{fig:overview}). These
high-gradient structures are mostly lost in the adaptive Gaussian smoothing.
By using the wavelet decomposition instead of smoothing, the shape of these
brightness features are better preserved.

We extracted wavelet components from the 0.8--4~keV image binned to 1\as\
pixels, using 6.6, 13, 26, 53 and 105 kpc scale wavelets, selecting the
thresholds of statistical significance in order to achieve balance between
retaining small-scale details and minimizing noise and artifacts. The wavelet
decomposition coefficients calculated for the 0.8--4~keV image were used for
all narrow-band images and their corresponding error images (that is, the same
smoothing was applied in all energy bands, as in \S\ref{sec:gausstmap}). Point
sources in the 6.6~kpc wavelet component were removed from those images before
coadding different scales. The resulting temperature map is shown in
\autoref{fig:tmap}{\em b}; it reveals the small-scale structure much better
than the one in panel {\em a}\/ at the expense of being only qualitative.

\section{Results}
\label{sec:results}

\subsection{Shock front (or fronts?)}

The bow shock to the SW of the cluster center, first reported in
\cite{2005ApJ...627..733M}, is readily apparent in the 0.8--4~keV image
(\autoref{fig:overview}) and in the temperature map (\autoref{fig:tmap}). The
latter shows a region of about 5~keV in front of the shock and 9--10~keV
behind the shock. We extracted spectra from 4 sectors in 2 annular regions in
front of the shock (S3-S10), and 3 sectors (S0-S2) including the shock
(\autoref{fig:spec}). In the pre-shock region, temperatures are $\sim 5\text{
keV}$ and are remarkably similar over this large area. Overall it appears that
pre-shock region is cool and undisturbed, with temperature falling with radius
slightly from $T=5.7\pm0.8$~keV (S3-S6 combined, $r\sim 650$ kpc from the
cluster center) to $T=4.5\pm0.8$~keV (S7-S10 combined, $r\sim 900$ kpc).
Behind the shock, in regions S0-S2, the temperatures span 8--14~keV. The
values are consistent with \cite{2005ApJ...627..733M} analysis of a shorter
dataset, who found $T=4.8^{+1.2}_{-0.8}\text{ keV}$ in front of the shock and
$T=11.5^{+6.7}_{-3.1}\text{ keV}$ behind (the latter value is deprojected,
therefore not directly comparable to that here).  In region S2, a cool blob of
gas appears to be projected onto the shock.  This feature is coincident with a
small but discernible brightness enhancement in the soft-band image. It could
be a splash or a broken off blob of the cool core inside the shocked gas.
Regardless of its origin, it may need to be masked in the analysis of the
shock, which will be the subject of a separate paper.

There is a kink in the shock surface (marked in \autoref{fig:overview}),
behind which (downstream from the shock) is a region of enhanced X-ray
brightness (``plume'' in \autoref{fig:overview}). The gas in the plume (region
S0) is as hot as the post-shock gas elsewhere, though the temperature map
(\autoref{fig:tmap}) suggests a mixture of different temperatures there. It
appears that a local gas inflow from the south is crossing the shock at that
location.

There is an apparent steepening of the surface brightness profile along the
NE-SW merger direction, northeast of the cluster center (located between
splashes B and C in \autoref{fig:overview}{\em b}) that looks like a
counterpart (``reverse'') shock to the main shock front. However, we do not
detect a significant difference in projected temperature between regions C3
and C4 (\autoref{fig:spec}) ahead and behind that brightness feature. The
presence of other features (splash B, bump C, the tail) makes this a crowded
location compared to the clean SW bow shock, and it is unlikely we can
deproject the emission correctly.

\subsection{Break up of the cool core remnant}
\label{sec:remnant}

\begin{figure*}
        \centering
        \leavevmode
        \includegraphics*[width=180mm]{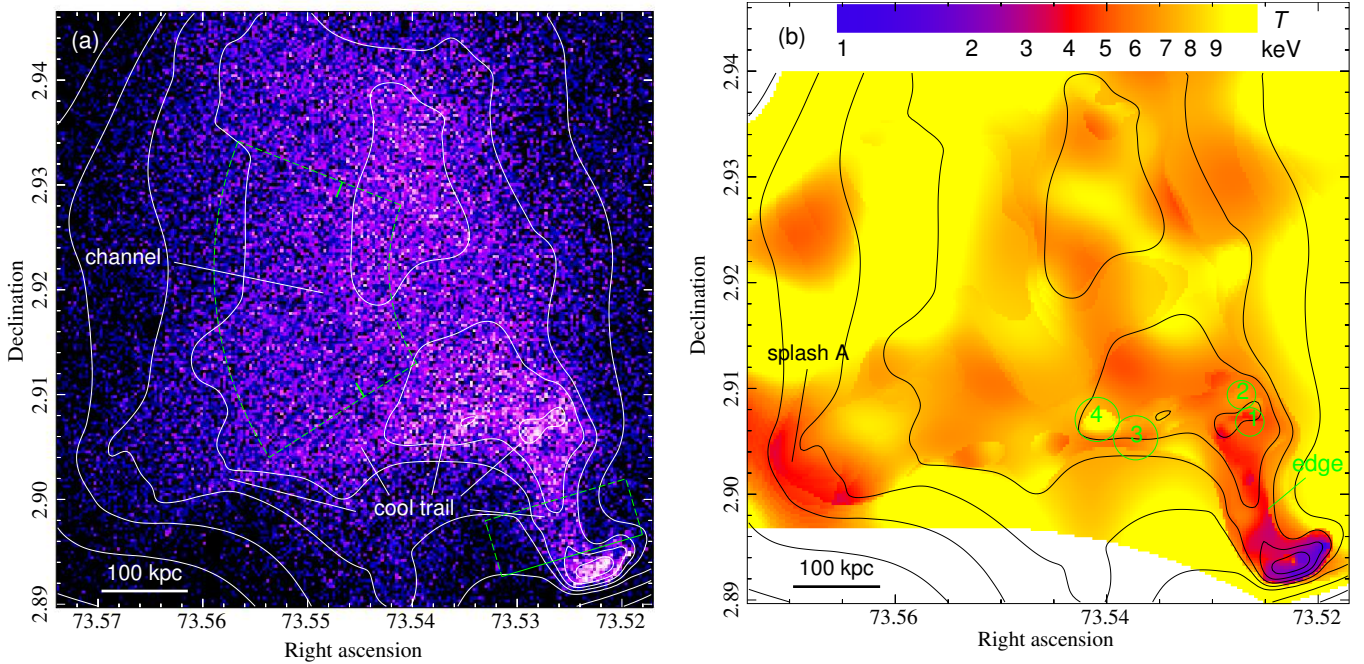}

        \caption{Zooming in on the wavelet temperature map of the stripped
          cool core remnant. ({\em a}) 0.8--4~keV image binned to 1\as\
          pixels. A radial profile within the dashed annular sector is shown
          in \autoref{sec:filament}, \autoref{fig:channel}. A profile in the
          rectangular region across the cool trail just above the foot is
          shown in \autoref{sec:conductivity}, \autoref{fig:edge}. ({\em b})
          similar to \autoref{fig:tmap}{\em b}, but derived without the
          largest wavelet scale 210~kpc. Overlaid are X-ray contours. Note the
          color scale is different from that in \autoref{fig:tmap}. The green
          labels are related to our discussion of thermal conductivity in
          \autoref{sec:conductivity}.}

        \label{fig:leg}
\end{figure*}

\begin{figure*}
        \centering
        \leavevmode
        \includegraphics*[width=180mm]{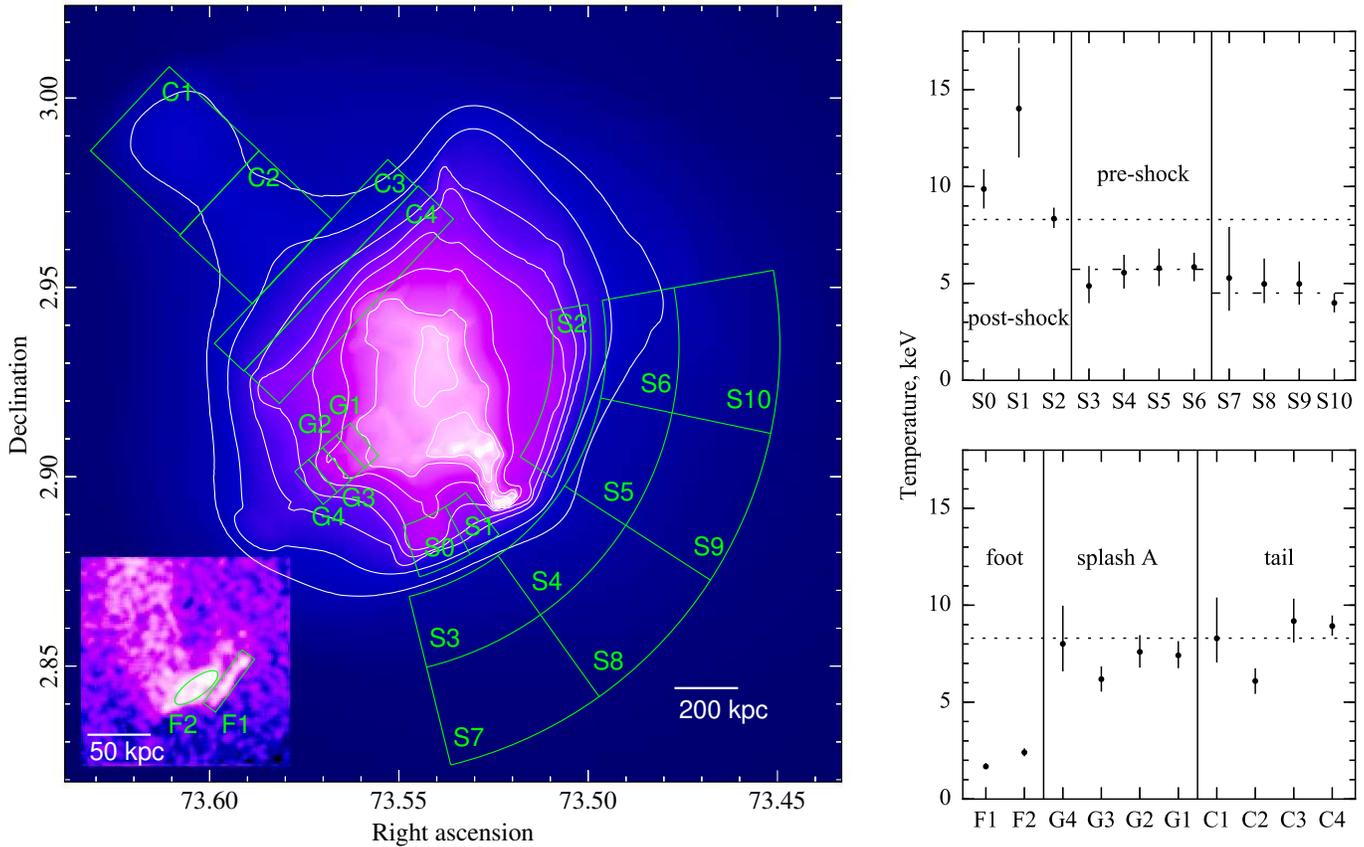}

        \caption{Spectral fitting regions shown on wavelet reconstructed
          0.8--4~keV image with inset showing $4\times$ enlarged view of
          slightly smoothed 0.8--4~keV image of the tip. Projected
          temperatures fitted in XSPEC are shown with 1-$\sigma$ error bars.
          The dotted line shows the cluster average temperature, and the
          dot-dashed lines for S3-S6 and S7-S10 show the best fit
          temperatures in those regions combined. }

        \label{fig:spec}
\end{figure*}

Behind the shock is a twisted structure resembling a leg (labelled in
\autoref{fig:overview}{\em b} and \autoref{fig:leg}). There are dense clumps,
as inferred from their high surface brightness, at the foot and at the knee,
and more along the ridge extending east from the knee (most pronounced in the
unsharp-masked image of \autoref{fig:overview}{\em d}). The foot (zoomed onto
in inset of \autoref{fig:spec}) is particularly striking. It consists of two
bright, very elongated ($50\times10$ kpc and $50\times 20$ kpc in projection)
clumps separated by a gap with an X-ray brightness contrast of $>2$. Their
projected temperatures are 1.5--2.5~keV (\autoref{fig:spec}); the narrower
finger on the outside is the colder of the two. There is no apparent galaxy
coincident with the foot, but the fingers are displaced from the center of the
BCG of one of the infalling subclusters by only $16\arcsec = 50$ kpc.

%%% FIGURE 3

%%% FIGURE 4

The wavelet temperature map in \autoref{fig:leg} shows that cool clumps trace
the structure extending north from the foot to the knee, which then turns
east, continuing toward ``splash A'' and ``splash D'' (\autoref{fig:tmap};
splashes will be discussed in \autoref{sec:splashes}). At the knee, a small
X-ray brightness cavity does not show a significant deviation in projected
temperature from the bright blobs above it. Not all the surface brightness
enhancements correspond to cool spots (as one would expect if the structure
were in pressure equilibrium), suggesting that projection effects are
significant.

The overall picture strongly suggests that the ``foot'' and the bent ``leg''
formed as a result of the disruption of a cool core, once hosted by the
subcluster centered on the BCG that is now ahead of the foot
(\autoref{fig:overview}). The cool core have been swept off its host by strong
ram pressure of the merger, but has not yet been completely mixed with the hot
surrounding gas. This is similar to the cool ``bullet'' in the Bullet cluster
displaced from the former subcluster host \citep{2002ApJ...567L..27M,
2006ApJ...648L.109C}, but, while the cool core in the Bullet cluster remains a
coherent shuttlecock structure, in A520 the disruption has gone much further.

To see if this picture is consistent with the properties of the cool clumps,
we estimate the gas specific entropy and check if it is similar to that in
typical undisturbed cool cores. We calculate the specific entropy using the
following definition (widely used in X-ray cluster work):
\begin{equation}
K=T n_e^{-2/3}
\label{eq:entropy}
\end{equation}
where $T$ is the gas temperature and $n_e$ is the electron number density. In
all of our analysis, we assume $n_e=1.17 n_{\rm H}$. Since the regions in
question are small and bright, they dominate the emission along the l.o.s., so
no deprojection is needed for a qualitative estimate. 

For the outer, thinner finger (F1 in \autoref{fig:spec}), $T=1.7\text{ keV}$.
If we use the size of the spectral fitting region and assume an elongated
shape, i.e. $10\times10\times50$~kpc square cuboid, the derived density is
$n_{\rm H}=2\times10^{-2}\text{ cm}^{-3}$, giving $K \approx 20\text{ keV
cm}^2$. Since the emission is actually confined to a narrower part of the
fitting region region, if we assume instead a cylinder of the same length
50~kpc and diameter of 5~kpc (half the width of the extraction region), the
density estimate increases by a factor $\sqrt{16/\pi}$ to $5\times10^{-2}
\text{ cm}^{-3}$, which gives $K \approx 12 \text{ keV cm}^2$. Alternatively,
if the blob is cap-like, taking the geometry of a flat cylinder 50~kpc in
diameter and 5~kpc thick, the density changes by a factor $\sqrt{8/5\pi}$ to
$1.4\times10^{-2}\text{ cm}^{-3}$, which gives $K
\approx 25 \text{ keV cm}^2$.

For the inner, wider finger (F2 in \autoref{fig:spec}), $T=2.4\text{ keV}$ in
an elliptical spectral extraction region. Its density is $n_{\rm
H}=1.3\times10^{-2}\text{ cm}^{-3}$, $K \approx 40\text{ keV cm}^2$ assuming
constant density for a prolate spheroid with symmetry axis in the sky plane,
or $n_{\rm H}=8\times10^{-3}\text{ cm}^{-3}$, $K \approx 60\text{ keV cm}^2$
for an oblate spheroid instead.

The entropy estimates vary by a factor of 2 for the different geometries
(elongated vs. cap-like) but are not drastically different. Since the specific
entropy could only have increased in the process of merger disruption (e.g.,
via mild shock heating), such specific entropy values, along with the high gas
densities, put these gas clumps confidently in the parameter space of the
central core regions of cool-core clusters where typically $K\sim15\text{
keV cm}^2$ as opposed to non-cool-core clusters where $K
\sim 150\text{ keV cm}^2$ \citep{2009ApJS..182...12C}. Thus, the cold gas
``leg'' indeed appears to be a trail of pieces of a merger-disrupted cool core
being swept by the gas flows. We will use this conclusion in
\autoref{sec:conductivity}.

\subsection{Splashes, bumps and islands}
\label{sec:splashes}

The eastward extension of the leg curves to the SE after about 300 kpc, and
ends with a steep brightness drop (``splash A'' in
\autoref{fig:overview}{\em b}) not far beyond. The gas at the dense side of
the brightness drop appears to be cooler than the surroundings, including the
gas along this structure but closer to the center. While the projected
temperature in region G3 (which contains the tip of the splash) is only
marginally lower than in regions G2, G1 in \autoref{fig:spec}, and the
temperature in region G4 in front of the splash is poorly constrained, the
temperature contrast becomes quite pronounced in the wavelet temperature map
in \autoref{fig:tmap}{\em b}. This splash looks like a hydrodynamic feature
caused by ``ram pressure slingshot'' \citep{2004ApJ...610L..81H}, in which a
rapid decline of ram pressure in a merger causes a parcel of gas to move into
the less-dense gas and expand adiabatically, forming a cool spot. In this
case, it could be one of the low-entropy clumps remaining of the cool core and
forming the cool leg.

North of the cluster center there is another hydrodynamic structure of likely
similar origin (``splash B'' in \autoref{fig:overview}{\em b}; also region T5
in \autoref{fig:tmap}). The surface brightness structure is picked out by
wavelet decomposition, which can be seen in the original image to appear like
a pointed stream of gas. The temperature maps indicate that this region is
cool. The gas there is not necessarily related to the cool core.

There is a subtle brightness island extending further SE from splash A, marked
``island D'' in \autoref{fig:overview} and \autoref{fig:tmap}, whose origin is
unclear. Its projected temperature is not well constrained but does not rule
out a cool structure.

Another splash-like structure (``bump C'' in \autoref{fig:overview}{\em b}) is
located symmetrically opposite splash B about the merger axis. Unlike splashes
A and B and island D, it coincides with one of the weak-lensing mass clumps.
Its projected temperature is in line with the cluster average and may even be
higher (as suggested by the wavelet map). This bump may have an entirely
different origin, a subcluster adiabatically accreting gas, similar to the
feature that we will discuss in
\autoref{sec:clump}.

\section{Discussion}
\label{sec:discussion}

\subsection{Scene of a `train wreck'}
\label{sec:wreck}

The detail-rich \chandra\ X-ray image and gas temperature maps of A520,
especially the map in which we subtracted the large-scale cluster emission
using wavelet transformation, tell a complex story about the events in this
merging cluster. From the X-ray and weak lensing data, we see a major merger
proceeding mostly along the NE-SW axis. The NE chain of subclusters have
apparently moved away from the collision site, completely stripped of their
gas and currently hosting only low-level bumps of X-ray emission (we will
discuss this in detail in \autoref{sec:clump}). The SW subcluster is also moving
away from the cluster center, driving a prominent shock front. Apparently,
this subcluster had a cool core, which is now being stripped by ram pressure,
leaving a trail of cool clumps --- ``foot'', ``knee'' and ``leg''. The
meandering shape of this trail, its ending with splashes A and D, together
with several other signs of complex hydrodynamics such as the kink in the
shock surface, the ``plume'' next to it and ``splash B''
(\autoref{fig:overview}), suggest a secondary collision along the north-south
direction. A curious X-ray ``channel'', possibly resulting from this secondary
merger, will be discussed in \autoref{sec:filament}. The full history and
details of this ``train wreck'' of a cluster may be understood better with a
dedicated hydrodynamic simulation. However, already our present broad-brush
understanding of the A520 merger lets us make three measurements that are
interesting from the cluster physics viewpoint.

\subsection{X-ray channel}
\label{sec:filament}

\begin{figure}
        \centering
        \includegraphics*[width=88mm]{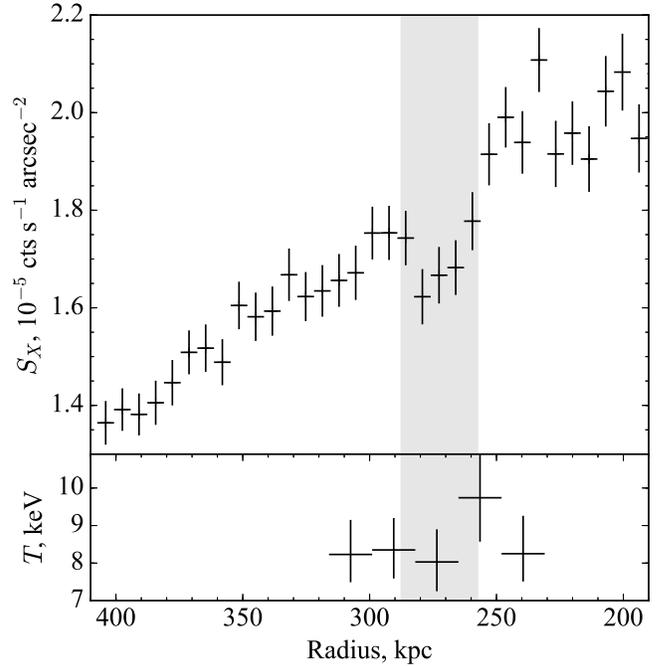}

        \caption{Radial profiles extracted in the annular sector in
          \autoref{fig:leg}{\em a}, of X-ray surface brightness (upper panel)
          and gas temperature (lower panel). The grey band is 30~kpc wide
          centered on the location of the channel, marked by white ticks in
          the profile extraction sector. Error bars for X-ray brightness and
          temperature are $1\sigma$. Radial distance is from the center of
          curvature of the sector. }

        \label{fig:channel}
\end{figure}

A close look at the X-ray image (in particular, \autoref{fig:leg}{\em a},
which show the image with different bin sizes, and \autoref{fig:overview}{\em
d}, which shows an ``unsharp-masked'' image), reveals a subtle, long X-ray
brightness ``channel''. It aligns with the direction of the secondary merger
that we mentioned above, running from the ``plume'' in the south through the
central region of the cluster toward ``splash B'' in the north
(\autoref{fig:overview}). We selected a sector in which this channel is most
apparent and which excludes any interfering features such as the leg, as shown
in \autoref{fig:leg}{\em a}. An X-ray brightness profile across the channel
extracted in this sector is shown in \autoref{fig:channel}. It confirms a
highly significant $\sim 10-12$\% drop in X-ray surface brightness. The
channel is about 30~kpc (9\as) wide and at least 200 kpc long, which is its
length within our sector, though the channel clearly extends beyond it and can
be traced as an X-ray dip in the leg and plume in the south, and similarly
further to the north.

The channel has to be a relatively thin sheet of lower-density gas seen along
the edge. If we assume a rough spherical symmetry of the main cluster body,
and assume that the channel is completely devoid of gas in 3D, the sheet's
extent along the l.o.s.\ would have to be $\sim 75$~kpc to give the observed
projected X-ray brightness drop. Since it cannot be completely empty, the
extent should be significantly greater.

It is interesting to speculate on the origin of the X-ray channel. First, we
note that X-ray ``cavities'' filled with radio emission are routinely observed
in cluster cool cores \citep[e.g.][and later works]{2000ApJ...534L.135M,
2002MNRAS.331..369F}; they are created by outbursts of the central AGN, where
the ejected relativistic matter expands and pushes the thermal gas away.

However, the channel/filament in A520 is not in a cool core, and its
$500-700$~kpc size is far greater than any of the cavities seen in cluster
cores. In principle, if in a certain region the magnetic field pressure
reaches levels comparable to the thermal pressure of the ICM, it may push
the plasma away from this region, in a manner similar to ``plasma depletion
layers'' observed near planets \citep[e.g.,][]{2004SSRv..111..185O} and
features seen in the galaxy cluster context in MHD simulations by
\cite{2011ApJ...743...16Z} (see their Fig.\ 23). Such a phenomenon may have
recently been observed by \cite{2016MNRAS.455..846W} in the core of the
Virgo cluster (though they observed X-ray enhancements rather than depletion
regions).

In such a scenario, the sum of thermal and magnetic pressure inside the
channel would equal the thermal pressure outside (assuming the magnetic
pressure outside to be negligible, as expected for the bulk of the ICM).
Neglecting projection effects --- that is, assuming the channel to be a broad
sheet spanning the whole cluster along the l.o.s. --- the observed drop in
X-ray brightness would correspond to a drop in gas density by $5-6$\% and a
drop in thermal pressure by $5-15$\% depending on the temperature behavior.
Such a drop of thermal pressure would imply a plasma $\beta_p$ parameter
($\beta_p\equiv p_{\rm thermal}/p_B$) reaching 10--20, compared to the usual
$\beta_p\sim {\rm few}\times100$. In a high-$B$\/ filament seen in simulations
by \citeauthor{2011ApJ...743...16Z}, both density and temperature of the gas
decline by similar factors, so the temperature is likely to decline in this
scenario.

Alternatively, the channel may be a purely hydrodynamic feature --- for
example, a region of shock-heated gas currently in thermal pressure
equilibrium, which has been squeezed into a sheet by gas flows. In this case,
the temperature in the channel should be higher by at least 5\% than that on
the outside.

To test these two possibilities, we extracted a projected temperature profile
in the same sector across the channel (\autoref{fig:channel}). It does not
show any significant temperature changes from the regions outside the channel,
but a 10\% deviation in either direction cannot be excluded. Thus,
both possibilities are viable on the basis of the X-ray data. If the channel's
span along the l.o.s.\ is less than assumed above, the 3D density and
temperature contrast may be higher (and the magnetic field in the first
scenario higher, too), but the projected surrounding denser gas would still
make it difficult to detect any temperature difference.

Both of the above configurations may have emerged as a result of a minor
merger along the north-south direction. For example, a small subcluster
infalling from the south (to explain the kink in the shock surface) and
crossing the main cluster could have stretched the magnetic fields in its
wake, and/or generated a shock-heated region. Subsequently, this region could
have been squeezed into a sheet --- for example, by large-scale gas motions of
the main NE-SW merger. One can also think of a radio-filled X-ray cavity swept
off one of the merging cluster cores, stretched by a N-S merger and compressed
into a sheet. It is unclear where that subcluster is now in the lensing mass
map (it may be clump N in \citealt{2008PASJ...60..345O}, which is not,
however, a particularly significant feature in \citealt{2012ApJ...758..128C}),
or how a low-density, unstable gas sheet could have survived as a coherent
structure in the middle of an ongoing merger. Such details might be clarified
by a dedicated hydrodynamic simulation. In all of the above scenarios, we
expect the magnetic field in the channel to be enhanced and oriented
preferentially along the channel (because of stretching and compression). This
may produce a bright filament in the cluster's giant radio halo
\citep{2001A&A...376..803G, 2014A&A...561A..52V}, because the synchrotron radio
emissivity is proportional to $B^2$, and that filament would be polarized.
Giant radio halos are unpolarized \citep{2012A&ARv..20...54F}, so this would
be a notable feature. The currently available radio data lack angular
resolution to test this prediction (Wang, Giacintucci, \& Markevitch 2016, in
prep.).

%%% FIGURE 5

%%%%%%%%%%%%%%%%%%%%%%%%%%%%%%%%%
\subsection{Dark subclusters in the northeast}
\label{sec:clump}

A520 exhibits a low X-ray brightness, relatively narrow tail, a subtle feature
but clearly visible out to about 1.3~Mpc northeast from the cluster center
(\autoref{fig:overview}; seen more clearly in a heavily-binned image in
\autoref{fig:clump}).  It has two broad X-ray peaks, each of which coincides
with a mass clump seen in the weak lensing map (\autoref{fig:clump}). The tail
and the clumps are aligned in the NE-SW direction of the main merger. The
outermost clump, centered 1.2~Mpc from the cluster center and approximately
0.5~Mpc in diameter, is particularly interesting, because it is relatively
free of projection of the rest of the messy cluster, which lets us make
several quantitative measurements.

Only two \chandra\ pointings (ObsIDs 9425, 9526) captured the tail, for an
effective exposure of about 200~ks. Spectra extracted from regions C1 and C2,
which approximately include the outer and inner of the two tail clumps,
respectively, show that they are both hot, with the outer tail clump (C1)
being slightly hotter than the inner (\autoref{fig:spec}).

The tail mass peaks are visible in two independent datasets, Subaru
\citep[see Fig.\ 11 in][]{2008PASJ...60..345O} and Magellan
\citep{2012ApJ...758..128C}. In the latter paper, only the inner tail peak is
shown (peak 1 in their Fig.\ 2); the outer, less significant peak is not shown
because it was outside the HST FOV, but it is seen in the uncropped version of
the map provided by D.\ Clowe, which we show in \autoref{fig:overview} and
\autoref{fig:clump}. The Subaru map covers a bigger field than Magellan or
\chandra\ and reveals another clump (their clump NE1) still further to the
northeast, but the Subaru map does not resolve these two Magellan tail clumps,
showing them as one (NE2).  For the quite substantial mass of the tail clumps
suggested by lensing, not much gas can be seen in the \chandra\ image, and not
much galaxy light is seen in the Subaru $i'$-band image either --- in
particular in the outer tail clump (Fig.\ 11c in
\citeauthor{2008PASJ...60..345O}).  This is very interesting in view of the
debated ``dark core'' in the center of A520; these clumps may be even
``darker'' and we will try to quantify this below.

We will now concentrate on the outer tail clump, because it is least affected
by X-ray projection. (The inner tail clump is more significant in the lensing
map, but it is hopeless to deproject it in X-rays.) We will compare the
specific entropy of the gas in the clump with that for the main cluster gas at
the same distance from the cluster center, estimate the clump total mass under
the hydrostatic equilibrium assumption, and derive a gas-to-mass ratio for the
clump.

\subsubsection{Specific entropy of the clump}
\label{sec:entro}

\begin{figure}
        \centering
        \leavevmode
        \includegraphics*[height=85mm]{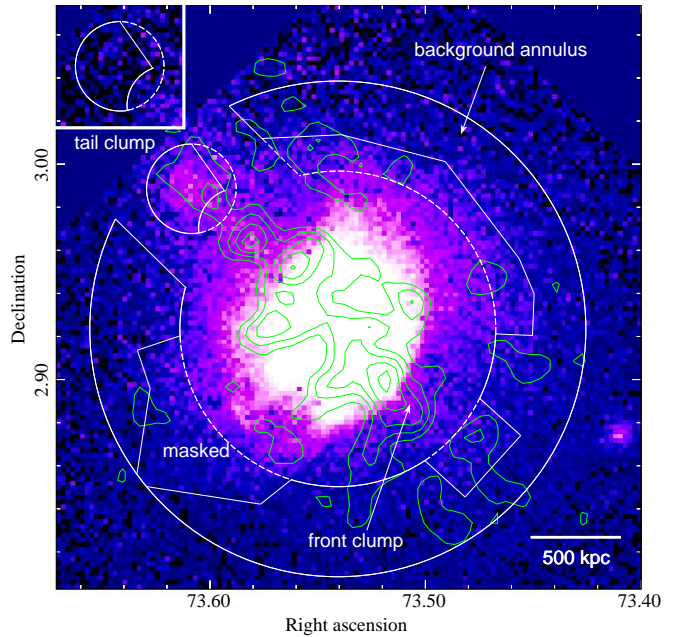}
        \caption{ The tail clump and background regions shown on
          the 0.8--4~keV image binned to 8\as\ pixels. Regions bounded by
          dashed lines were masked. Contours are lensing mass from D. Clowe.
          The top left inset shows the residual image after subtracting the
          clump and outskirt models (same color scale).}

        \label{fig:clump}
\end{figure}

To derive the gas density, we fit the heavily-binned X-ray image
(\autoref{fig:clump}) with a simple model consisting of two
spherically-symmetric 3D $\beta$-model density profiles --- one for the clump
and another for the main cluster outskirt near the radius of the clump. The
$\beta$-model profile is given by
\begin{equation}
n_{\rm H}(r)=n_{\rm H,0} \left[ 1 + \left(\frac{r}{r_c} \right)^2
\right]^{-3\beta/2}
\label{eq:betamodel}
\end{equation}
where $r_c$, $n_{\rm H,0}$ and $\beta$ are free parameters. Integrating $n_{\rm H}^2$\/
along the l.o.s. gives an observed X-ray surface brightness profile
(more precisely, the projected emission measure, which is very close to the
surface brightness for the relevant range of gas temperatures and the
\chandra\ energy band) in the form
\begin{equation}
\Sigma_X(\theta) \propto \left[ 1 + 
                \left( \frac{d_A \theta}{r_c}\right)^2
                \right]^{-3\beta+1/2},
\label{eq:2dbetamodel}
\end{equation}
where $d_A$ is the angular diameter distance and $\theta$ the angular distance
from center.

For the cluster outskirt, we extracted a 0.8--4~keV radial surface brightness
profile in an annulus around the same distance from the cluster center as the
clump, with prominent asymmetric features (tail including the clump, foot,
shock, splashes) masked out as shown in \autoref{fig:clump}. It is not
obvious where the ``center'' of a messy merger is; for this exercise, the
center is selected as a centroid of the X-ray emission at the relevant radii
in the outskirts. We fit the profile in this annulus using a model given by
\autoref{eq:2dbetamodel}, fixing the core radius $r_c$ at a typical value of
180~kpc (since we fit very far from the core). To determine the normalization
$n_{\rm H,0}$, we extracted a spectrum in the same region, fit it in XSPEC
using APEC model, and compared the model emission measure integrated over the
region $\int n_{\rm H} n_e dV$ with the absolute APEC model normalization
given by XSPEC. The best-fit projected temperature is
$T=4.1_{-0.9}^{+1.4}$~keV, and the beta-model parameters are $\beta =
0.62_{-0.05}^{+0.04}$ and
$n_{\rm H,0} = (4.4^{+1.2}_{-1.0}) \times 10^{-3}$~cm$^{-3}$.
At the clump's radius, the outskirt density is 
$n_{\rm H}=(1.3\pm0.1)\times10^{-4}$~cm$^{-3}$
(density in the outskirt is better constrained than the beta-model
normalization, which is an extrapolation of the profile in the outskirt).

%%% FIGURE 6

The clump density model was then fitted in 2D (that is, pixel-by-pixel, as
opposed to extracting a radial profile), because the cluster outskirt
contribution makes the surface brightness distribution non-radial. We added a
$\beta$-model density component for the clump to the density model for the
outskirt, fixing the latter at its best fit derived above (which masked out
the clump region with a good margin). We chose to add the clump density
component, rather than replacing one with the other in the 3D region of the
clump, to avoid any smoothness issues for the hydrostatic mass estimates; this
choice does not matter as long as the model fits the X-ray image well. The sum
of the two density components was calculated in 3D and a projected emission
measure was calculated for each pixel of the X-ray image in a masked
near-circular region of $r=250$~kpc (\autoref{fig:clump}). The best-fit shape
parameters for the clump are $\beta = 0.80\pm0.07$ and $r_c =
203^{+20}_{-16}$~kpc (uncertainties determined with the other parameter fixed
at best-fit value) and the model fits the image well ($\chi^2=135/199=0.68$).

To derive the absolute gas density in the clump, we need the gas temperature.
If we assume the clump to be isothermal with the outskirt, its density
normalization can be derived directly from the X-ray surface brightness and
the outskirt model derived above. This gives a density of
$n_{\rm H}=(1.0\pm0.1)\times10^{-3}$~cm$^{-3}$
at the clump, of which the clump component dominates the outskirt component by
a factor of 7 --- a significant gas overdensity.

However, the clump appears to have a higher projected temperature than the
outskirt, $T=8.1_{-1.9}^{+3.6}$~keV (for region C1 in \autoref{fig:spec},
which covers the clump), and its 3D temperature should be higher still.
Therefore, we also consider the case in which an isothermal, but hotter, clump
is embedded in the outskirt. We make a simple assumption that all gas within a
$r=250$~kpc sphere of the clump is at a higher temperature. We generate a
model image with a cutout for this sphere and calculate the projected
contribution of the 4~keV outskirts to the clump spectrum (it is about 9\% in
projected emission measure at the center of the clump). Adding this as a
``background'' model for the spectrum of the clump, we obtain a
``deprojected'' clump temperature 
$T=9.7_{-3.3}^{+5.5}$~keV,
which is slightly
higher but consistent with the projected temperature (as expected, given the
relatively high brightness contrast) and the density at the center of the
clump increased by 10\% to 
$n_{\rm H} \approx 1.1\times10^{-3}$~cm$^{-3}$
 compared to
the isothermal assumption --- a negligible change for our qualitative
estimates, and considering the systematic uncertainties due to the unknown
geometry.

Using the deprojected temperature and density for the clump, we can estimate
the specific entropy of the gas at its center, defined as in
\autoref{eq:entropy}, $K=930_{-320}^{+510}$~keV~cm$^2$ (error accounts only
for the uncertainty in temperature). For comparison, the gas in the outskirts
has $K=1540^{+530}_{-340}$~keV~cm$^2$ at this radius. The two values are
consistent, and both are consistent with the entropy range $(1-2)
\times 10^3$~keV~cm$^2$ observed at $r\sim 1-1.3$~Mpc for a large sample of
clusters \citep{2009ApJS..182...12C}. The temperature and density of the gas
in the clump are consistent with adiabatic compression of the 4~keV gas from
the outskirts perturbed by the gravitational attraction of the clump.  In
contrast, for cool cores, \citeauthor{2009ApJS..182...12C} observe $K<
50$~keV~cm$^2$, far below the observed value for the clump, so this gas cannot
be a remnant of a former cool core (like the ``foot'', \autoref{sec:remnant}).
We will speculate on the sequence of events that could have created this clump
after estimating its mass below.

\subsubsection{Total mass of the ``dark clump'' and its possible origin}
\label{sec:mass}

Given the relative isolation of the tail clump, we can try to estimate its
total mass, assuming that its hot gas is close to hydrostatic equilibrium with
the clump's gravitational potential. The equilibrium should be achieved on a
timescale of sound crossing the size of the subcluster. Considering that the
subcluster is unlikely to move supersonically at such a distance from the core
(we also do not see any shocks around it), this assumption should be adequate
for a qualitative estimate.

The total enclosed mass within the radius $r$\/ for a spherical mass clump is
given by \citep[e.g.,][]{1988xrec.book.....S}
\begin{equation}
M(<r) = - \frac{k T(r) r}{G \mu m_p}\left[\frac{d \ln n_{\rm H}}{d \ln r} +\frac{d
  \ln T}{d \ln r}\right],
\end{equation}
where $\mu$ is the mean atomic mass per gas particle ($\mu \approx 0.6$ for
ICM), $T(r)$\/ is the local gas temperature at the radius $r$\/ and $n_{\rm
H}$ is the gas density, which is the sum of the clump and outskirt density
models in our case. For an accurate estimate, a temperature profile is
required, for which our data are not adequate --- all we know is that the
temperature near the clump center is around 10~keV and it goes down to 4~keV
in the main cluster's outskirts. Therefore, we will make two isothermal
estimates for these two temperature values to get a rough range of masses.
(The higher-temperature estimate would neglect the $(d \ln T/d \ln r)$
contribution, which should be nonzero in this case, partially canceling out
the effect of the expected lower local $T$\/ at the radius of the estimate.)
For the gas density gradient, we will use the best-fit model (sum of offset 3D
beta-models) obtained above, calculating the gradient in the direction
tangential to the main cluster in order to isolate the effect of the clump. We
will calculate the mass for a radius well within our model fit above. Within a
$r=200$~kpc sphere, we obtain the total mass of $2.5\times 10^{13}$~\msun\ and
$6\times 10^{13}$~\msun\ for the lower and higher temperature values,
respectively (of course, statistical errors do not matter with such a modeling
uncertainty). This is consistent with masses within the same radius derived
for real mid-temperature clusters \citep[e.g.,][]{2006ApJ...640..691V}.

To assess the sensitivity of the clump hydrostatic mass estimate to our
assumption of spherical symmetry for the main cluster's outskirt, we varied
the surface brightness of the outskirt by factor $\pm2$ in the region of the
clump and refitted the density model for the clump. The resulting variations
in the quantity $d \log n_{\rm H}/d \log r$ (where $n_{\rm H}$ is the sum of
the clump and outskirt components, and $r$ is the distance from the center of
the clump), which determines the clump mass estimate, varies by at most 40\%
in the radial range of interest. Thus, our estimate should be relatively
robust to geometric assumptions.

It is interesting to compare our mass estimate with a weak lensing mass for
this clump. D. Clowe (private communication) provided us with an estimate of a
projected mass within a cylinder of $r=150$~kpc. Depending on whether the HST
data (partially covering the clump) are included in the reconstruction along
with the Magellan data, the projected mass is $(1.7-2.3)\times 10^{13}$~\msun;
the statistical significance of this clump detection is only 2--3$\sigma$. To
convert our 3D measurement into a projected mass, we assume that the clump's
total mass profile is truncated at $r=300$ kpc. For the low and high
temperatures, we obtain the projected masses within the $r=150$~kpc aperture
of $2.4\times10^{13}$~\msun and $5.6\times 10^{13}$~\msun, respectively. The
lower range of our X-ray estimates is in agreement with the lensing value.

With this qualitative validation for our mass estimate, we now estimate the
gas mass fraction $f_{\rm gas}$ for the clump. Within the $r=200$~kpc sphere,
we get $f_{\rm gas}=0.03$ and 0.014 for the cool and hot clump assumptions,
respectively. This is low --- even the former, conservatively high value is at
least a factor 2 below the $f_{\rm gas}$ values observed within the same radius
in relaxed clusters \citep[e.g.,][]{2006ApJ...640..691V}. So the tail clump
appears to be ``dark'' in terms of the apparent deficit of both the galaxy
light and the ICM density. The caveat here is the uncertainty in the total
mass is quite high, and one cannot be entirely confident in the X-ray
hydrostatic equilibrium assumption here; a more sensitive weak lensing
observation may reduce the total mass and $f_{\rm gas}$ uncertainty.

Based on the high specific entropy that we derived in \autoref{sec:entro}
(consistent with that in the A520 outskirts), a cluster-like total mass and an
anomalously low gas fraction, we speculate that this clump entered the
collision site from the SW as a fairly massive subcluster. It then lost all of
its gas to ram pressure stripping (and probably all matter in its outskirts to
tidal stripping) during the passage through the main cluster, but re-accreted
some high-entropy gas from the A520 outskirt once it emerged on the other
side. The gas compressed adiabatically into its potential well once the
subcluster slowed down sufficiently. Of course, the resulting $f_{\rm gas}$
need not be anywhere near the universal value. On subsequent infall, such a
subcluster would be the analog of the dark-matter dominated ``gasless''
subclusters used in idealized hydrodynamic simulations
\citep[e.g.,][]{2006ApJ...650..102A, 2010ApJ...717..908Z}, which disturb the
gravitational potential but produce few hydrodynamic effects.

Judging from the X-ray/lensing overlay, the more prominent inner-tail lensing
mass peak \citep[clump 1 in][]{2012ApJ...758..128C} appears to have a similar
or even lower gas-to-mass ratio (the peak X-ray brightness is similar and the
lensing mass is higher). We did not attempt any quantitative X-ray estimates
for this clump because the 3D geometry is very uncertain.

We also note that the mass clump that hosted the stripped cool core, denoted
``front clump'' in \autoref{fig:clump}, appears to be reaccreting or
concentrating the surrounding hotter gas. It is seen as an enhancement in
density of the preshock gas at the position of the clump. Although this
subcluster appears to be more massive than the tail clump, its gas density
enhancement is smaller, probably because the gas is flowing over this dip in
the gravitational potential toward the shock front with a higher velocity. As
this subcluster moves to the periphery and slows down with respect to the gas,
it may re-accrete a gas halo similar to that of the tail clump.

Interestingly, \cite{2015ApJ...806..123S} observed three massive weak-lensing
subhalos in the periphery of the Coma cluster with {\it Suzaku}. One of their
subhalos exhibits a diffuse X-ray emission excess with the projected gas
temperature similar to that of the surrounding ICM. They derive an extremely
low gas fraction of $\sim0.001$ for it. These subhalos may be of similar
nature to our dark clump --- complete stripping of the original gas and
subsequent reaccretion of the surrounding ICM.

\subsection{Constraints on thermal conduction}
\label{sec:conductivity}

Thermal conductivity is one of the important but poorly known properties of
the ICM. It is determined by the topology of the tangled magnetic field frozen
into the ICM and by plasma microphysics. The heat transport should be
completely suppressed across the field lines (because the electron gyroradii
are many orders of magnitude smaller than other relevant linear scales in the
ICM, \citealt{1988xrec.book.....S}), while heat may flow along the lines
between those regions of the ICM for which such a path along the lines exists.
However, even {\em along}\/ the filed lines, the heat transport may be
strongly suppressed in a high-$\beta_P$ plasma (such as the ICM) because of
micro-scale plasma instabilities \citep[e.g.,][]{2008PhRvL.100h1301S}.

Observationally, few definitive measurements have been done. Across cold
fronts, where the temperature jumps abruptly, thermal conductivity has been
shown to be suppressed by at least a couple of orders of magnitude compared to
the Spitzer value \citep[][and later works]{2000MNRAS.317L..57E}.  The likely
explanation is the magnetic field ``draping'' along the cold front surface as
a result of the gas flowing around it, which effectively isolates the two
sides of the front from each other \citep{2001ApJ...549L..47V,
2007PhR...443....1M, 2011ApJ...743...16Z}. Some constraints outside the
special regions of cold fronts have been reported, based on the existence of
spatial temperature variations in the ICM. For example,
\cite{2003ApJ...586L..19M} derived an order of magnitude suppression of
conductivity between regions of different temperature in the body of a hot
merging cluster A754, and \cite{2012A&A...541A..57E} derived a large
suppression factor based on the survival of a tail of cool gas stripped from a
group infalling into the hot cluster A2142. In both cases, the physical
significance of the constraints is ambiguous because the topology of the
magnetic fields is unclear --- for example, it is possible (and in the case of
the infalling group, likely) that the observed regions of the different
temperature come from separate subclusters whose magnetic field structures
remained topologically disconnected even after the merger, so there are simply
no pathways for heat exchange (as suggested in \citealt{2003ApJ...586L..19M}).
Indirect upper limits on the effective isotropic conduction based on the
analysis of ICM density fluctuations have also been derived
\citep[e.g.,][]{2013A&A...559A..78G}.

In our picture of A520, the cool clumps in the ``leg'' (from the ``foot'' to
the ``knee'', then east along the bright ridge) come from the same cool core
(\autoref{sec:remnant}), so their magnetic field structure should be (a)
interconnected and (b) stretched along the trail by the same gas motions that
separated the cool core pieces. This offers a unique opportunity to constrain
the conductivity {\em along}\/ the field lines. We know the Mach number of the
shock front and the velocity of the post-shock flow
\citep{2005ApJ...627..733M}, which lets us estimate how long ago they were
stripped based on their distance along the trail. We can then determine if the
conductivity between them should be suppressed by comparing the Spitzer
conduction timescale with their age,
\begin{equation}
\kappa / \kappa_S = (t_{age} / t_{cond} )^{-1}.
\end{equation}

In our simple picture, the ``foot'' is the last piece of the former cool core
that is still gravitationally bound to the subcluster that drives the shock
(or, at least, it has been bound until recently). The post-shock gas flow
peels away pieces of the cool core, carrying them off at the downstream
velocity of 1000~km~s$^{-1}$ \citep{2005ApJ...627..733M}. Guided by the
temperature map (\autoref{fig:leg}{\em b}), we picked two pairs of circular
regions in near contact (in projection) that have large and significant
temperature differences. The blobs are assumed to attain their present
temperature and spatial separation upon stripping from the core, and then to
move with the flow together; the distance of the pair from the ``foot'' along
the ``leg'' gives the age of the pair.

We estimated the thermal conduction timescale as in, e.g.,
\citet{2003ApJ...586L..19M}:
\begin{align}
t_{\text{cond}} \approx 1.2\times10^7 & 
   \left(\frac{n_e}{2\times10^{-3} \text{ cm}^{-3}} \right) \nonumber \\
   & \left( \frac{l_T}{100 \text{ kpc}} \right)^2 
     \left( \frac{T}{10 \text{ keV}} \right)^{-5/2} \text{ yr,}
\label{eq:Mcond}
\end{align}
where $n_e$ is the electron number density, $l_T \equiv T/|\nabla T|$ is the
thermal gradient scale length, and $T$\/ is the electron temperature. This
equation applies when the heat flux is unsaturated --- where $l_T \gg
\lambda_e$, the electron mean free path \citep{1956pfig.book.....S}:
\begin{equation}
\lambda_e \approx 31 \text{ kpc } 
    \left( \frac{kT}{10 \text{keV}}\right)^2 
    \left( \frac{n_e}{10^{-3} \text{ cm}^{-3}} \right)^{-1}.
\label{eq:mfp}
\end{equation}
The regions we selected are far from saturation. The density in \autoref{eq:Mcond}
is taken to be the average density in the corresponding stretch of the leg,
$n_{\rm H}=0.01$~cm$^{-3}$. This is uncertain to a factor 2, based on density
estimates for each region using two different geometric assumptions
--- all emission originating from a sphere in projection (leading to higher
densities and therefore longer $t_{cond}$), or from cylinder along the l.o.s.
that is 400~kpc long (the opposite effect). Therefore our values of
$\kappa_S/\kappa$ also has a factor of 2 uncertainty arising from this.

We also consider how the uncertainty in $l_T$ affect our results.  Since
$t_{cond} \sim l_T^2$, it is important to estimate the gradient correctly.
For one set of estimates, we use the projected temperatures in the regions of
interest, measured using XSPEC. However, projection is likely to wash out the
temperature gradient, resulting in longer $l_T$. While our wavelet temperature
map (\autoref{sec:wavtmap}) is qualitative, it removes most of the projection
effects and leaves only the relevant linear scales. \autoref{fig:leg}{\em
b}\/ shows a temperature map created with only the smallest wavelet components
that correspond to the angular scale of the structures in the leg. Using the
temperature values from this map, the values of $T/\Delta T$ are up to 2 times
smaller. We note that since we calculate the gradients using projected
distances between the regions, this is a lower limit for $l_T$. On the other
hand, the leg may be bent along the l.o.s., so our ages for the region pairs
may be underestimated. And of course, the absence of a temperature gradient
does not always result from thermal conduction, so we can only place a lower
limit for an order-of-magnitude estimate of a suppression factor.

The results are shown in \autoref{table:conduction}. For regions 1 and 2 (see
\autoref{fig:leg}{\em b}), we cannot say whether the conduction is suppressed
--- the suppression factor is consistent with 1 for both the projected or
deprojected temperatures. For regions 3 and 4, $\kappa_S/\kappa
\sim 3.3-11$, so there seems to be some suppression.

\begin{table}
  \begin{center}
    \caption{Thermal conduction timescale estimates. The columns are:
    estimated age of the feature in yr; projected temperatures in~keV;
    suppression factor ($\kappa_S/\kappa=t_{\rm age}/t_{\rm cond}$) using
    projected temperatures; deprojected temperatures from the wavelet
    temperature map; suppression factor using deprojected temperatures.}
    \label{table:conduction}
    \begin{tabular}{cccccccc}
        \hline\hline
        Reg  & $t_{\rm age}$, yr & $T_1^{\rm proj}$ & $T_2^{\rm proj}$  &
        $\kappa_S/\kappa^{\rm proj}$ & $T_1^{\rm dep}$ 
        & $T_2^{\rm dep}$ & $\kappa_S/\kappa^{\rm dep}$ \\
        \hline
        1, 2  & $1.9\times 10^8$  & 5.2  &  7.4 & 1.1 & 4.5 & 7 & 1.4 \\
        3, 4  & $2.6\times 10^8$ & 7.4 &  11.9 & 3.3 & 6 & 14 & 11 \\
        \hline
    \end{tabular}
  \end{center}
\end{table}

We did not use splash A at the end of the cool trail for this estimate, even
though there appears to be a significant temperature gradient there. The
splash should have been cooling via adiabatic expansion as it formed, so its
age is very uncertain.

\begin{figure}
        \centering
        \leavevmode
        \includegraphics*[width=85mm]{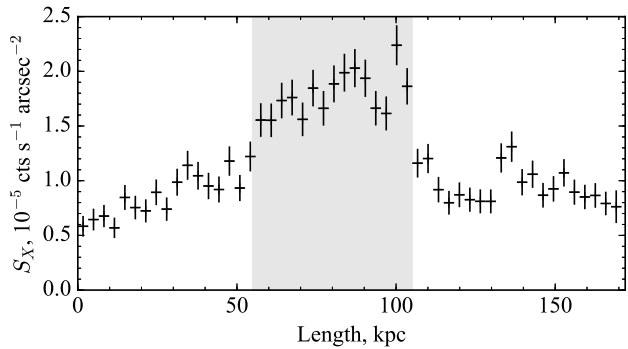}
        \caption{ Surface brightness profile across the cool trail just above
        the foot, extracted in a narrow rectangular region
        (\autoref{fig:leg}{\em a}), showing an unresolved edge at around 105
        kpc. The cool trail spans the shaded region between 55 and 105 kpc.
        The small bump between 130 and 150 kpc is due to the tip of the foot.}

        \label{fig:edge}
\end{figure}

The above attempted constraints for the suppression {\em along}\/ the field
lines can be contrasted with thermal conductivity across the edge of the cool
trail of gas. In our scheme for A520, the cool trail should be isolated from
the surrounding gas by a magnetic field stretched along its boundary (a likely
analog of the infalling group in \citealt{2012A&A...541A..57E}). For example,
consider the feature marked `edge' in \autoref{fig:leg}{\em b}. Along this
trail of cool gas the temperature gradient is small, but in the perpendicular
direction it jumps from about 4.5~keV in the leg to 12~keV for the post-shock
gas on a scale smaller than 10~kpc. The surface brightness jump there is
unresolved by \chandra\ (\autoref{fig:edge}). The trail is 120~kpc long,
implying an age of $1.2\times10^8$~yr from the cool core at the downstream
velocity. The density inside the trail is estimated from the emission measure
in the same region (assuming cylindrical shape) to be
$6\times10^{-3}$~cm$^{-3}$. For these values, $\lambda_e=3.5$~kpc, so this is
still in the unsaturated conduction regime. We find $t_{cond}=7\times10^5$~yr,
implying $(\kappa/\kappa_S)^{-1}\gtrsim 170$. Thus, this trail could not have
formed in the presence of any significant thermal conduction across the edge.

%%% FIGURE 7

\section{Summary}
\label{sec:summary}

The deep \chandra\ exposure of Abell 520 revealed rich structure in this
cluster train wreck, including a prominent bow shock. Some of these
structures provide interesting constraints on cluster physics. We derived
detailed gas temperature maps using two methods, one that utilizes
variable-width smoothing and evaluates the projected temperature, and
another that uses wavelet decomposition to ``deproject'' the large-scale
structure in a qualitative way and enhance the contrast of the interesting
small-scale structure.

On small scales, A520 exhibits an apparent disrupted cool core at a unique
evolutionary stage --- the gas of the core is swept away from the central
galaxy of its former host subcluster by ram pressure of the gas flow
downstream of the shock front, completely displacing the gas peak from the
galaxy (by 50--70 kpc).  The disrupted core is not mixed with the hot gas
but still forms a physically connected trail of dense clumps (a cool
``leg''). Its twisted structure apparently reflects the chaotic gas
velocities in this region. The core remnant in A520 is at a later stage of
disruption compared to the bullet in the Bullet cluster, where it is still
seen as a regular shuttlecock structure. The specific entropy of the gas in
the clumps is much lower than elsewhere in the cluster and is typical of
other cool cores. 

In the above scenario, the magnetic field within the leg should be stretched
along the leg and still connect the clumps (since they come from the same
core), while insulating the leg from the surrounding hot gas. We use the
observed temperature variations between the cool leg and the surrounding
gas, and within the leg, to constrain thermal conductivity across the field
lines (a factor $>100$ suppression from the Spitzer value) and, for the
first time, suggest that the conductivity along the lines may also be
suppressed by a factor of at least several. This is, of course, dependent on
our assumption about the magnetic field structure.

About 1.3 Mpc northeast of the cluster center, the X-ray image reveals a
subtle tail of low X-ray brightness. Two clumps in the tail coincide with
mass peaks seen in the weak lensing mass map. For one of the clumps that is
least affected by projection, we derived a specific entropy of the X-ray
gas, which turns out to be similar to the high value for the cluster gas at
that radius, while the gas density in the clump is several times higher.
Thus, the X-ray enhancement at that clump appears to be due to adiabatic
compression of the surrounding gas. The second clump looks similar, though
quantitative estimates are difficult because of projection.  It appears that
these clumps have passed through the cluster merger site and lost all of
their gas (or, alternatively, arrived to the cluster already gasless) and
then re-accreted the surrounding outskirt gas as soon as they slowed down
sufficiently. An X-ray hydrostatic estimate the total mass of the clump is
consistent with the lensing mass. The ratio of the X-ray measured gas mass
to total mass is 1.5--3\%, much lower than the typical average cluster
value, making these clumps truly ``dark subclusters''. Of course,
considering our scenario for their origin with stripping and re-accretion, it
would have to be a coincidence if the resulting gas fraction ended up the
same as the universal cluster value.

Finally, we found a curious long ($>200$ kpc), narrow (30 kpc or 9\as) X-ray
``channel'', going across the bright cluster region along the direction of
an apparent secondary merger. The projected X-ray brightness in the channel
is 10--12\% lower than in the adjacent regions. The channel has to be a
sheet spanning at least 75 kpc along the l.o.s. It is possible that this is
a ``plasma depletion layer'' with the magnetic field stretched and enhanced
by the merger; the plasma $\beta$\/ parameter should reach 10--20 in the
sheet. In this scenario, we predict that the channel will be seen as a
bright filament in the radio image of sufficient angular resolution, and the
filament will be polarized.

The prominent bow shock in this cluster will be studied in our subsequent
work (Wang, Giacintucci, \& Markevitch 2016, in prep.).

\vspace{5mm}

We thank the referee for useful comments that made the paper clearer. QW was
supported by Chandra grants GO3-14144Z and GO5-16147Z.

\bibliography{cluster}

\end{document}